\documentclass[english,aps,pra,reprint,noshowpacs,superscriptaddress]{revtex4-1}   
\usepackage[T1]{fontenc}	
\usepackage{geometry}		
\usepackage{graphicx}
\usepackage[above,below]{placeins}	
\usepackage{times}
\usepackage{hyperref}  
\hypersetup{colorlinks=true,urlcolor=blue,citecolor=blue,linkcolor=blue}   
\begin{document}

\preprint{APS/123-QED}

\title{The difficulties associated with integrating computation into undergraduate physics.}

\author{Ashleigh Leary}
\affiliation{Department of Physics and Astronomy, Michigan State University, East Lansing, MI 48824}
\author{Paul W. Irving}
\affiliation{Department of Physics and Astronomy, Michigan State University, East Lansing, MI 48824}
\author{Marcos D. Caballero}
\affiliation{Department of Physics and Astronomy, Michigan State University, East Lansing, MI 48824}
\affiliation{CREATE for STEM Institute, Michigan State University, East Lansing, MI 48824}
\affiliation{Department of Physics \& Center for Computing in Science Education, University of Oslo, N-0316 Oslo, Norway}


\begin{abstract}
\verb+Abstract.+ 
From a department being resistant to change to students not buying into the new computational activities, the challenges that are faced with integrating computation into the physics undergraduate curriculum are varied. The Partnership for Integration of Computation into Undergraduate Physics (PICUP) aims to expand the role of computation in the undergraduate physics curriculum. The research presented in this paper is part of a larger project examining the role of the PICUP workshop in facilitating both the integration of computation into classrooms and developing a supportive community to support this integration. An important part of providing the necessary supports for integration is understanding and categorizing the problems members of this community of integrators face when integrating computation in their courses. Through individual and group interviews, we discuss the barriers to integration that new and experienced community members of PICUP have experienced in the past or perceive could exist in the future.
\end{abstract}

\maketitle


\section{Introduction}
The integration of computation into the undergraduate physics curriculum is becoming an important priority for the community of undergraduate physics educators \cite{Chonacky2008IntegratingDevelopments}. There is a large pool of people who teach undergraduate physics who are shifting toward utilizing computation in some sense in their physics classes \cite{Caballero2017}. This shift is being motivated in some part by the NGSS standards, which are promoting this idea of helping the next generation of scientists develop computational thinking \cite{NGSSLeadStates2013NextPractices,Cooper2015ChallengeLearning,AAPTUndergraduateCurriculumTaskForce2016AAPTCurriculum}. Those standards, designed with K-12 education in mind, are influencing instructors at the college level as well \cite{Irving2017P3:Environment}. However, as with any change in teaching practice, there are difficulties and challenges to be negotiated in order to integrate successfully. 
%
%
%
%

The need for facilitation of this negotiation sparked the formation of The Partnership for Integration of Computation into Undergraduate Physics (PICUP). The purpose of PICUP is to create a dialogue for those interested in integrating computation and those who have already completed it successfully. This dialogue consists of pedagogy, methodology, and issues that are important when it comes to integrating computation into the undergraduate physics curricula. The PICUP group also uses this dialogue to accomplish their overall goal of getting undergraduates comfortable and competent with computation. PICUP realizes that computation is the third arm of physics, alongside theory and experiment. They also realize that students who are not learning this skill are at a disadvantage in this increasingly data-rich and model-driven society. 

The PICUP Faculty Development Workshop (FDW) is a week-long event that is focused on introducing faculty to curriculum and pedagogical ideas in the hope that the attendees
will then successfully implement those ideas when they
return to their home institutions \cite{Irving2017UnderstandingPractice}. The workshop provides an opportunity for attendees to engage with implementers who hold expertise and experience in implementing computation into their curriculum. During the workshop there is no explicit discussion of the challenges and difficulties one may face when integrating computation but the PICUP community places an emphasis on providing post-workshop support that is aimed at providing some support in this area. But despite these efforts, the members of the PICUP community still lack knowledge around integrating computation smoothly and effectively. 

\section{Literature Review}

There has been prior research about curriculum reform when integrating or reforming teaching methods using approaches based on Physics Education Research (PER) \cite{Henderson2005TeachingProfessors}. The work completed by Dancy and Henderson on the constraints of implementing research-informed practices plays an important role in examining the challenges of teaching physics in general \cite{Dancy2005BeyondPractices}. They highlight the conundrum of the thriving nature of PER as a research field that produces a number of results that indicate the effectiveness of PER-informed approaches while at the same time highlighting that there is no widespread adoption of PER-informed practices. In fact, many faculty in physics are familiar with PER practices but for a number of reasons do not implement the suggested recommendations. 

Dancy and Henderson examined this problem by interviewing faculty members and asking them about their around PER based approaches \cite{Henderson2005TeachingProfessors}. Their research identified that the main issue lies with inconsistency between the beliefs of the instructors and the actual enacted practice of those beliefs. The reasons for these inconsistencies fall into five main systemic forces: student resistance, time structure, departmental norms, expectations of content coverage, and lack of instructor time. The fact that they are called ``systemic forces'' is telling. Dancy and Henderson make sure to acknowledge that these instructors do not act in isolation. Rather, they are in the system of their institution. That system can choose to support or resist PER-informed practices, and the unfortunate reality is that the systems that instructors find themselves in typically resist. In the following paragraphs the five systemic forces from Dancy and Henderson \cite{Dancy2005BeyondPractices} are discussed in order to highlight replicated results in the interviews we had with faculty.

\textbf{Student Resistance:} Students resist by not supporting research-based methods. Dancy and Henderson specify this by stating that there's a norm that has been built for these students. They do not have to be willing participants and instead can expect to go to class and not engage with activities if they choose not to.  PER-based approaches often encourage making the classroom more interactive for them and initial resistances by the students is common place.


\textbf{Time Structure:} This refers to the fact that it is not possible to change how many weeks there will be of class. Semester structures at institutions are not flexible and of a fixed length of time, which does not allow for individual differences in learning needs. Combine this with students who are taking multiple other courses and that limits the amount of time they can spend on one class. 

\textbf{Departmental Norms:} Every department also has norms, and these norms can either help or hinder an instructor that wants to change the curriculum. Especially when the idea for the change comes from PER-informed results, the norms for a department could be highly traditional and that makes instigating change a larger challenge. 

\textbf {Expectations of Content Coverage:} This refers to the often experienced problem of having too much material that needs to be covered. When there is too much material to be covered, instructors will dismiss research-based methods that are geared toward developing a deep understanding in order to get through all of the material. 

\textbf {Lack of Instructor Time:} Lastly, instructors have a lot of responsibilities on their plate. With all that is expected of them, there is not always time to learn and integrate new techniques which is an obvious barrier to them implementing new approaches.

The prevalence of any of these systemic forces is what creates the inconsistencies between what the instructors want to teach and what actually happens in the classroom. It is unclear that the challenges that are highlighted in this previous research will be replicated in the context of computation integration or if there is striking differences in the challenges one will face. 


\section{Methods}
Semi-structured interviews were conducted with thirty individuals, sixteen of which were in focus groups. All of the interviewees were considered in some way to be a part of the PICUP community, but the level at which they influence the community varies. The level at which they teach physics also varies with attendees teaching at the university level, the community college, and even one teacher from a high school. All of these instructors were given gendered pseudonyms. These interviews were part of a larger project examining the role of the PICUP FDW in the development and expansion of a community of practice focused on integrating computation into the undergraduate curriculum. Therefore the focus of the discussion was on the PICUP community, its supports, and the individuals' roles within that community. However, a proportion of the interview was also focused on the challenges that the interviewees had faced both historically in integrating computation and the challenges they were facing at the time of the interview. Each interview was transcribed and then analyzed using a thematic analysis approach with a focus on both preconceived and emergent themes. Themes were then compared and contrasted, then combined and described when similarities existed.

The interviews took place before, during, and after the PICUP FDW. This fact, along with the differing levels of physics that these instructors teach, make for a diverse set of experiences. For example, Anakin is a leader within the PICUP community. He is part of the movement to change the undergraduate physics curriculum with the addition of computation, and wants to make it easy for new instructors to do this. When he started integrating computation on his own, he had no support from his department and was practically a 'lone wolf'. His main challenge was with his department. Then there is Beru. She was someone who participated in the PICUP FDW, and her interview takes place after the workshop. She has the freedom to implement whatever she wants into her classes, but gets stuck on the little details of how to design the activities. Her main challenge is not in dealing with the department like it was for Anakin, but knowing what to do with the support that she has. The experiences that Anakin and Beru have are very different, and so the problems that they highlight when asked what barriers they came across are different. The differences are not so large that they cannot be categorized generally, but they are different enough to need a context more specific than just the Dancy/Henderson paper.

\section{Results}

The initial set of theme's presented highlight the replication of the systematic forces that Dancy and Henderson discovered but filtered through the lens of the integration of computation being the approach that is being resisted. 

\subsection{Replicated results in computation context}

\textbf{Student Resistance:} When dealing with computation students resist in a different fashion. They are still rejecting research-based methods in general, but they are not specifically rejecting working in groups or being forced to think.  

\begin{quote}
\textit{"...my experience with them [has] also been that sometimes it's almost like asking them to eat vegetables, and they don't like it, you know?" - Han}
\end{quote}

Instead, they are rejecting learning something new, and activities they think do not belong in a physics class. In many instances, instructors aired frustrations that they perceived they had students who were perfectly capable of accomplishing a computational task but simply refused to because of their own opinions of it.

\textbf{Time Structure:} The time issue replicates almost explicitly when adding computation as unfortunately, a semester at an institution is, and for the most part always will be, the same fixed length.

\begin{quote}
\textit{"If I could devote the first three weeks to have the students really understand the answer to the question what is a rate...if they spent three weeks using Euler to understand what is a rate, that's three weeks well spent." - Wedge}
\end{quote}

Adding computation to a curriculum creates a new challenge of teaching the skill and the lesson that is encompassed in that skill at the same time. Basically, syntax or how to use a Jupyter Notebook needs to be taught in addition to the physics of what the program is teaching. This can be quite challenging when there is a lot to teach and not a lot of time. 

\textbf {Departmental norms:} This is different because of the types of norms that exist when discussing computation. 

\begin{quote}
%
%
%
\textit{"The other thing is that some of the theorists I think have expressed ...sort of second hand...some concern of courses adopting computation to the detriment of like what they feel...physics classes should be, which is a paper and pencil calculation." - Lando}
\end{quote}

Instructors consistently mentioned that they had to focus a lot on the language they used around faculty members at their institutions when it came to bringing up integrating computation. They had to be careful that they did not frame adding computation in a way that would make it seem more important than what the faculty was already doing. 

\textbf {Expectations of Content Coverage:} was not as prevalent in the context of computation. It was not a question of what computational material should be covered, it was more an issue of time. 

\begin{quote}
\textit{"I got a lot of push back from the TAs because they were saying, 'Oh yeah, okay, this is neat, but our students still don't know how to add vectors.' Or, 'This is neat, but we have an exam coming up in two weeks and we ought to be spending this time studying what's going to be on the exam.'" - Wedge}
\end{quote}

There are concepts in computation that need to be covered and there was not any expectations that needed to be met within computation. It was more a question of should computation be added and a different analytical lesson taken away. There seemed to be underlying expectations of different material that needed to be covered as opposed to computation itself. 

\textbf {Time of Instructors:} The time of instructors is still valuable and the things that preoccupy them are still the same (e.g. large teaching loads and research). 

\begin{quote}
"...[I have other goals] right now too...Chief of which is you know getting tenure. So while being an agent of change to include computation more broadly could be something that enhances my tenure profile. I'm not quite sure how I would do that right now. So I'm more focused on some of my other efforts." - Dodonna
\end{quote}

Integrating computation asks even more from instructors as they may already have materials from previous semesters to use to teach a class. It can go another step further if the instructor is asked, or wants to, design an entirely new course based on computation in physics. 

\subsection{New systemic forces for computation context}

\textbf{Lack of Instructor Knowledge:} There are plenty of instructors that want to integrate computation into their curriculum, but do not know how to start. This prevents them from being able to accomplish the goals they have for their students, and can cause other problems to appear (e.g., the department will not support them because they do not know what to do). This lack of knowledge can manifest in two different ways.  

The first way is when an instructor displayed that they had a lack of experience with coding themselves. This made them uncomfortable as they would be teaching their students how to do something that they were not an expert on. They could be lacking in experience with anything from Jupyter Notebooks to Python. The problem was that they could not begin to know what to educate the students on when they did not know what the fundamentals of coding were on their own.

\begin{quote}
\textit{"The only snag that I hit was I decided to experiment with GlowScript. And, and you know, my students are... getting comfortable with vPython, then I said, you know, you could use GlowScript...but  I wasn't too prepared for GlowScript myself. And so I just quickly backed off and went back to vPython." - Han }
\end{quote}

The second way is when the instructor knows how to code, and may even be a relative expert on it, but does not know how to design activities for students that are just learning. In addition to this, they may not know how to tie in valuable computational lessons to the regularly scheduled physics lessons that are found in most introductory classes. Assessing the level of student skill in  computation is  difficult as some students come to class with previous experience and others have never seen a piece of code before. Learning how to do design activities with this in mind takes a lot of trial and error, and having someone who has done is before can be a great asset. 

\begin{quote}
\textit{"I didn't need help so much with the technical language stuff, but like, you know, How should I ... You know, Should I break this up into multiple exercises? How long would this take? How much time should I dedicate to this? Should I have the ball without air resistance and the ball with air resistance animate at the same time, or one after the other?" - Poe}
\end{quote}

\textbf{Accessible Platform:} An issue that is specific to computation is that it requires a medium to be taught. This requirement hinders instructors by forcing them to evaluate what platform to use and if they can be supported using that platform.

Choosing a platform is a difficult task due to the fact that there is not a universal physics coding language that every professor uses regularly. There are pros and cons to each platform that have to be looked at, and each person has their own opinions on what the best thing to learn is. Additionally, picking a platform that is easily accessible and understandable for students is always something to be taken into consideration. 
\begin{quote}
\textit{"Because our departments, some of the people said, 'Well, we know how to use MATLAB, right, use this or that.' And it wasn't until one of the astronomers spoke up and says, ''Well, we're going towards using Python.' Only astronomers are. But some of the people were still saying, 'Well, why don't you keep that to a small part of your course because it's not really very important.'" - Wedge}
\end{quote}

The other problem is choosing a platform that is affordable and easy to access for both professors teaching a class and students alike. This is a problem that can be more common among high school teachers and community college professors. 

\begin{quote}
\textit{"So when I first started...I wanted to have access to VPython for all the students. And what I needed was a class set of laptops. I needed VPython installed on all of them, and it took ... It actually took two years to get the funding. To get the laptops." - Luke}
\end{quote}

\textbf{The IT Crowd}: Nothing can be more off putting to trying something new in your classroom than having to deal with the IT department.

\begin{quote}
\textit{"One thing I'm currently doing is fighting with our IT. Obviously when you want to do something new, first taking it up with IT and saying 'Okay, I have no experience how to do it. Can you help me?' That usually doesn't go very far, so that's why I implemented the original JupyterHub server on my own office work station. But we have a new work station policy...they want everything to be secure, secure, secure...they closed the ports on which I run my JupyterHub server." - Ackbar}
\end{quote}

The issue that Ackbar is highlighting is one that is unique to integrating computation. It should be surprising that it is not mentioned more times in the amount of interviews that were had, but maybe it has become so embedded in any work place that people just think is a part of life. 

\section{Discussion and Conclusion}
The overwhelming motivation for these instructors wanting to integrate computation was typical, "I wish I had been taught this when I was an undergrad," or, "It's something students are going to encounter and I want them to be prepared." It is concerning how nearly every interviewee was able to identify at least one issue they ran into when trying to accomplish something that they perceive as being beneficial for their students. However, it should not be surprising to anyone that this is the case -- water is wet of course.  

The first step in helping to solve any problem is identifying it. Hopefully, instructors who read this paper, will gain comfort in knowing that these barriers have happened to someone else and that PICUP is a whole collection of people who have faced these barriers and come out victorious, and they have advice to share with those only just jumping in.  The more we know and learn about the issues surrounding integrating computation the easier it will be for new instructors to join the thriving community of those teaching computation to their students. 

\bibliography{Mendeley_Ashleigh_PICUP}

\end{document}